\documentstyle[multicol,aps,prl]{revtex}

\begin{document}

\title{Bare Quark Matter Surfaces of Strange Stars and  $e^+e^-$
Emission}

\author{V.V. Usov}

\address{Department of Condensed Matter Physics, Weizmann Institute
of Science, Rehovot 76100, Israel}

\maketitle

\begin{abstract}
  
We show that the Coulomb barrier at the quark surface of a hot
strange star may be a powerful source of $e^+e^-$ pairs
which are created in an extremely strong electric field of the barrier
and flow away from the star.
The luminosity in the outflowing pair plasma depends on 
the surface temperature $T_{_{\rm S}}$ and may be very high, up to 
$\sim 3\times 10^{51}$ ergs s$^{-1}$ at $T_{_{\rm S}}\sim 10^{11}$ K. 
The effect of pair creation by the Coulomb barrier may be a good
observational signature of strange stars which can give an answer to
the question of whether a compact object is a neutron or strange
star.

\end{abstract}


\begin{multicols}{2}

With the discovery of pulsars and their identification with the neutron 
stars, many  people thought that neutron star matter is the
ground state of matter at high density. Later, this belief was
challenged by Witten \cite{W84}. He proposed that strange matter 
made of quarks is the ground state at ultra-high density. If the
Witten's hypothesis is true, then neutron stars with a sufficiently
high central density may transform themselves into strange stars
\cite{AFO86}. In this case, at least some part of the compact
objects known to astronomers as pulsars, powerful accreting X-ray
sources, X-ray bursters, soft $\gamma$-ray repeaters, etc. might 
be strange stars, not neutron stars as it is usually assumed 
\cite{AFO86,G90}.
In principle, the question of whether compact objects
are neutron or strange stars can be answered by
comparing the core density of a neutron star to the strange matter 
transition density. Unfortunately, all the nuclear-physics calculations
to date do not yield a definitive answer. As to available data on 
pulsars and other compact objects, they are not able to give us 
an answer either \cite{WSWG96}. This is because the bulk 
properties of models of strange and neutron stars of masses that 
are typical for neutron stars, $1\lesssim M/M_\odot \lesssim 1.8$,
are relatively similar. The situation changes, however, as regards the
possibility that
strange quark matter with the density of $\sim 4\times 10^{14}$
g cm$^{-3}$ may be, by hypothesis, the absolute ground state of the 
strong interaction (i.e., absolutely stable with respect to
$^{26}$Fe) and can exist up to the surface of strange stars. This differs 
qualitatively from the case of the neutron star surface and
opens a unique possibility to observe a superdense quark matter.

"Normal" matter (ions and electrons) may be at the quark surface of
strange stars. The ions in the inner layer are supported against 
the gravitational attraction to the underlying strange star 
by a very strong electric field of the Coulomb barrier in the quark 
surface vicinity (see below). There is an upper limit to the amount of 
normal matter at the quark surface, $\Delta M\lesssim 10^{-5}M_\odot$ 
\cite{GW92}. Such 
a massive envelope of normal matter with $\Delta M\sim 10^{-5}M_\odot$
completely obscures the quark surface. However, it was pointed out 
\cite{HPA} that a strange star at the moment of its formation is very 
hot. The temperature in the interior of such a star may be as high as a 
few $\times 10^{11}$ K. The rate of mass ejection from an envelope of 
such a hot strange star is very high \cite{WB92}, and if any normal 
matter remains at the quark surface of a strange star at $t\gg 10$ s,
its mass is many orders smaller than the maximum.
Besides, high temperatures lead to a considerable reduction of the 
Coulomb barrier, which favors the tunneling of nuclei toward the strange 
star surface \cite{KWWG}. Therefore, it is natural to expect that the 
quark surface of a very young strange star is nearly (or completely) 
bare. Recently, it was argued \cite{Usov97a} that the normal-matter 
atmosphere of such a star remains optically thin in spite of a high  
accretion of gas onto the star until the quark-surface temperature is 
higher than $\sim 3\times 10^7$ K. Such an atmosphere
does not obscure the quark surface and has no influence on the
surface structure. In this Letter, some properties 
of the bare quark surfaces of strange stars are considered.

At the bare surface of a strange star the density changes abruptly from 
$\sim 4\times 10^{14}$ g cm$^{-3}$ to zero.
The thickness of the quark surface is about 1 fm, which is a 
typical strong interaction length scale. There are some electrons in
quark matter to neutralize the electric charge. The density of electrons
is $\sim 10^3-10^4$ times smaller than the baryon density of the 
quark phase. The electrons, being bound to the quark matter by the
electromagnetic interaction and not by the strong force, are able
to move freely across the quark surface, but clearly cannot move to
infinity because of the bulk electrostatic attraction to the quarks.
The electron distribution extends up to $\sim 10^3$ fm above the quark
surface \cite{AFO86}. Associated with this electron layer 
is a strong electric field, which is radially outwardly
directed. This field prevents the electrons of the layer from
escaping to infinity. Introducing a local electron chemical 
potential $\mu_e(r)$, we have $\mu_e(r) - eV(r)={\rm const}= 0$,
where $V(r)$ is the electrostatic potential at the distance $r$
from the stellar center \cite{KWWG}. The last
equality follows from the fact that
at $r=\infty$ there are no electrons, $\mu_e(\infty )=0$, and $V(\infty )
=0$. If $kT_{_{\rm S}}$ is much less than the Fermi energy of electrons
$\varepsilon_{_{\rm F}}$, the temperature dependence of $\mu_e(r)$ is 
negligible, and in this case
we have $\mu_e(r)\simeq \varepsilon_{_{\rm F}}$. Taking into account that 
$\varepsilon_{_{\rm F}}\simeq 20$ MeV inside quark matter, the 
potential difference across the Coulomb barrier at the quark surface
is $\Delta V_{_{\rm C}}\simeq 2\times 10^7$ V.
The thickness of the region with a strong 
electric field is $\Delta r_{_E}\simeq 10^3$ fm $\simeq 10^{-10}$
cm at $kT_{_{\rm S}} \lesssim mc^2\simeq 0.5$ MeV and decreases with
increase of $T_{_{\rm S}}$, where $m$ is the electron mass \cite{KWWG}.
The typical magnitude of the electric field in this region is
$\sim 5\times 10^{17}$ V cm$^{-1}$ \cite{AFO86}. 
This field is a few ten
times more than a critical field  

\begin{equation}
E_{_{\rm cr}}=m^2c^3/e\hbar \simeq 1.3\times 10^{16}\,\,\,\,{\rm V\,cm}
^{-1}\,,
\label{Ecr}
\end{equation}

\noindent
where $\hbar$ is Plank's constant and $e$ is the electron charge. 

It is well known that a vacuum 
with such a strong electric field is unstable, and $e^+e^-$ 
pairs are created spontaneously. The pair 
creation rate per unit volume and per unit time in a stationary and
homogeneous electric field, ${\bf E}={\rm const}$, is 
\cite{Schwinger,Rein}

\begin{equation}
W_\pm ={m^4c^5\over 4\pi^3\hbar^4}\left({E\over E_{\rm cr}}
\right)^2\sum_{n=1}^{\infty}{1\over n^2}\exp\left[-\pi n\left(
{E_{\rm cr}\over E}\right)\right]\,.
\label{W1}
\end{equation}

\noindent
At $E\gg E_{\rm cr}$, we have 

\begin{equation}
W_\pm \simeq {m^4c^5\over 24\pi\hbar^4}\left({E\over E_{\rm cr}}
\right)^2\simeq 1.7\times 10^{50}\left({E\over E_{\rm cr}}
\right)^2\,\,\,\,{\rm cm}^{-3}\,{\rm s}^{-1}\,,
\label{W2}
\end{equation}

\noindent
In such a strong electric field the value of $W_\pm$ is very high, and,
at first sight, the Coulomb barrier has to be a very powerful source of 
$e^+e^-$ pairs, irrespective of the surface temperature.
But this is not the case. The matter is that, strictly speaking,
Eqs. (\ref{W1}) and (\ref{W2}) are valid only if no real 
particles are present. The electrons in the quark-surface vicinity
occupy some part of the energy levels and can significantly
reduce the rate of pair creation.

Let us use Dirac's hole picture for considering 
of pair creation by the Coulomb barrier. Figure 1 shows the space
dependence of the potential energy $eV(r)$ as well as the
corresponding energy gap of the Dirac equation between $mc^2 - eV(r)$
and $-mc^2 - eV(r)$. Pair creation results from the tunneling of an
electron from the "Dirac sea" through the classically forbidden zone.
The probability for such a tunnel process 
is described by a penetration factor given by \cite{Rein}

\begin{equation}
P\simeq \exp\left(-{2\over \hbar}\int^{r_+}_{r_-}q(r)dr\right)\,,
\label{P}
\end{equation}

\noindent
where $q(r)=\sqrt{m^2c^2-(\varepsilon -eEr)^2/c^2}$ is the imaginary
momentum and $\varepsilon$ is the energy of the tunnelling electron, the
electron rest energy $mc^2$ is included into $\varepsilon$. 

At $T_{_{\rm S}}=0$, we have $\mu_e=\varepsilon_{_{\rm F}}$, and
the tunnelling of an electron from the "Dirac sea" may be
possible only into the states with energies below the pair creation 
threshold, $\varepsilon \leq\varepsilon_{_{\rm F}} 
-2mc^2$ (see Fig. 1.). However, at zero temperature all electronic
states with energies $\varepsilon\leq \varepsilon_{_{\rm F}}$ 
are occupied. Hence, at $T_{_{\rm S}}=0$ such a tunnelling 
of electrons is kinematically forbidden, and there is no $e^+e^-$
pair creation by the Coulomb barrier as it was adopted by tacit consent 
in all studies of strange-quark matter (e.g., \cite{AFO86,KWWG}). 

At finite temperatures, $T>0$, a quantum-mechanical state with an energy 
$\varepsilon$ cannot definitely be said to be occupied or empty.
Instead an occupation probability function $f(\varepsilon )$ may be
used. For electrons in thermodynamical equilibrium,
Fermi-Dirac statistics requires $f(\varepsilon )$
to be of the form (e.g., Ref. \cite{Rein})

\begin{equation}
f(\varepsilon )=\left[1+\exp\left({\varepsilon-\mu_e\over kT}
\right)\right]^{-1}\,,
\label{f}
\end{equation}

\noindent
i.e. all electronic states are not completely occupied, $f(\varepsilon 
)<1$. In this case,
the tunnelling of electrons from the "Dirac sea" into the empty 
states is kinematically allowed. Therefore, the Coulomb barrier at the
quark surface may be a source of $e^+e^-$ pairs if the 
surface temperature is non-zero.

The thickness of the classically forbidden zone is (see Fig. 1.)

\begin{equation}
\Delta r_{\rm forb}=r_+-r_-={2\hbar\over mc}{E_{\rm cr}\over E}
\simeq 7.7\times 10^{-11}{E_{\rm cr}\over E}\,\,\,\,{\rm cm}\,.
\label{Dx}
\end{equation}

\noindent
In our case, $E\gg E_{\rm cr}$, we have $\Delta r_{\rm forb}\ll\Delta r_
{_{E}}=r_{\rm max}-r_{\rm min}$, 
and therefore, for the tunnel process the electric field 
of the Coulomb barrier may be considered as a homogeneous one. For such a 
field, Eq. (\ref{P}) yields 

\begin{equation}
P\simeq \exp [-\pi ({E_{\rm cr}/ E})]\,,
\label{P1}
\end{equation}

\noindent
i.e. the penetration
factor $P$ is exactly the exponential factor in Eq. (\ref{W1})
for the one-pair term, $n=1$. At $E\gg E_{\rm cr}$, from Eq. (\ref{P1})
we have $P\simeq 1$, and the electrostatic
barrier for the tunnel process completely disappears.
In this case, the rate of pair production when electrons are created 
into the empty states is extremely high,
and all the empty states below the pair 
creation threshold (see Fig. 1.) are occupied, $f(\varepsilon )=1$, by
creating electrons very fast. Then, the rate of pair creation by
the  Coulomb barrier is determined by the process of thermalization
of electrons which favors the empty-state production below the
pair creation threshold.

Positrons are created mainly at the outer boundary of the Coulomb 
barrier, i.e. at $r\simeq r_{\rm max}$, and then, they are ejected 
from the barrier by the electric field. Outflow of positrons results 
in a small decrease of the potential difference across the Coulomb 
barrier, which, in turn, results in outflow of electrons from the 
electron layer. The flux of $e^+e^-$ pairs from the 
quark surface is 

\begin{equation}
\dot N_\pm\simeq 4\pi R^2\Delta r_{_E}\Delta n_e t_{\rm th}^{-1}\,,
\label{Npm}
\end{equation}

\noindent
where $R$ is the radius of the star, $\Delta n_e$
is the density of electronic empty states with energies below the
pair creation threshold at thermodynamical equilibrium
and $t_{\rm th}$ is the 
characteristic time of thermalization of electrons.

At $kT_{_{\rm S}}\ll \varepsilon_{_{\rm F}}$, we have \cite{Usov97b}

\begin{equation}
\Delta n_e\simeq {3kT_{_{\rm S}}\over \varepsilon_{_{\rm F}}}
\exp \left(-{2mc^2\over kT_{_{\rm S}}}\right)n_e\,.
\label{Dne}
\end{equation}

\noindent  
where $n_e$ is the density of electrons.

In the electron layer, the spectrum of electrons
is thermalized due to electron-electron collisions, and
the thermalization time is of the order of $ \nu_{ee}^{-1}$, 
where \cite{Gena}  

\begin{equation}
\nu_{ee}\simeq {3\over 2\pi}\sqrt{\alpha\over \pi}
{(kT_{_{\rm S}})^2\over \hbar \varepsilon_{_{\rm F}}}J(\zeta )
\label{nuee}
\end{equation}

\noindent
is the frequency of electron-electron collisions for degenerate electrons
with $\varepsilon _{_{\rm F}}\gg mc^2$,
$\alpha = e^2/\hbar c=1/137$ is the fine structure constant,                                      

\begin{equation}                                                                J(\zeta )\simeq \cases{
51(1-19.5\zeta^{-1}+296\zeta ^{-2})\,\,\,\,
{\rm at}\,\,\zeta\geq 20\,,
\cr
\noalign{\vskip3pt}
0.23\zeta^{1.8}\,\,\,\,\,\,\,\,\,\,\,\,\,\,\,\,\,\,\,\,\,\,
\,\,\,\,\,\,\,\,\,\,\,\,\,\,\,\,\,\,
{\rm at}\,\,
1<\zeta<20\,,
\cr
\noalign{\vskip3pt}
(1/3)\zeta^3\ln (2/\zeta ) \,\,\,\,\,\,\,\,\,\,\,\,\,\,\,
\,\,\,\,\,\,\,\,\,\,\,\,\,\,\,\,\,
 {\rm at}\,\,\zeta \leq 1\,,
\cr}
\label{J}
\end{equation}

\begin{equation}
\zeta = 2\sqrt{{\alpha\over \pi}}{\varepsilon_{_{\rm F}}\over k
T_{_{\rm S}}}\simeq 0.1{\varepsilon_{_{\rm F}}\over k
T_{_{\rm S}}}\,.
\label{zeta}
\end{equation}

For $\varepsilon_{_{\rm F}}\simeq 20$ MeV, $R\simeq 10^6$ cm, 
$\Delta r_{_E}\simeq 10^{-10}$ cm and $n_e\simeq 10^{35}$ cm$^{-3}$,
from Eqs. (\ref{Npm}) -- (\ref{zeta}) the flux of $e^+e^-$ pairs from 
the bare quark surface of a strange star is

$$\dot N_\pm\simeq 3\times 10^{59}\left({k T_{_{\rm S}}\over 
\varepsilon_{_{\rm F}}}\right)^3\exp\left(-{2mc^2\over 
k T_{_{\rm S}}}\right)J(\zeta )\,\,\,\,{\rm s}^{-1}$$

$$\simeq \exp (-{1.2\times 10^{10}\,{\rm K}/ 
T_{_{\rm S}}})$$

\begin{equation}
\times
\cases{
10^{56}\ln(T_{_{\rm S}}/10^{10}\,{\rm K})\,\,\,\,{\rm s}^{-1}
\,\,\,\,\,\,{\rm at}\,\,\,T_{_{\rm S}}\gtrsim 
 10^{10.3}\,{\rm K} \,,
\cr
\noalign{\vskip3pt}
10^{55.5}(T_{_{\rm S}}/10^{10}\,{\rm K})^{1.2} \,\,
{\rm s}^{-1}\,\,\,
{\rm at}\,\,
10^9\,{\rm K}\lesssim T_{_{\rm S}} <10^{10.3}\,{\rm K}\,,
\cr
\noalign{\vskip3pt}
10^{57.3}(T_{_{\rm S}}/10^{10}\,{\rm K})^3 \,\,\,\,{\rm s}^{-1}
\,\,\,\,\, {\rm at}\,\,\,T_{_{\rm S}} <10^9\,{\rm K} 
\cr}
\label{Npm2}
\end{equation}

\noindent
within a factor 2-3 or so. The thermal energy of the star is a source of 
energy for the process of pair creation.

At high temperatures, $T_{_{\rm S}}\gtrsim 10^{10}$ K, the thickness,
$\Delta r_{_E}$, of the region with a strong electric field decreases with 
increase of $T_{_{\rm S}}$ \cite{KWWG}. On the other hand, the electron
density, $n_e$, increases with increase of $T_{_{\rm S}}$. As a result,
the value of $\Delta r_{_E}n_e$ is more or less independent
of $T_{_{\rm S}}$, and therefore, in Eq. (\ref{Npm2}) we ignored the 
temperature dependences of both $\Delta r_{_E}$ and $n_e$. 

The density of created $e^+e^-$ pairs at the stellar surface   
is $n_\pm\simeq \dot N_\pm/4\pi R^2c\lesssim 10^{33}$ cm$^{-3}$ that is
much smaller than the typical density of electrons in the electron layer,
$n_\pm\ll n_e$. Therefore, the production of $e^+e^-$ pairs by the 
Coulomb barrier does not affect essentially the structure of 
the barrier itself. Our consideration of the process of $e^+e^-$ pair           creation is valid only for strange stars with bare (or nearly bare)
quark surfaces.

At extremely high temperatures, $T_{_{\rm S}}\simeq$ a few $\times
10^{11}$ K, the flux of pairs $\dot N_\pm$ is as high 
as the upper limit

\begin{equation}
\dot N_\pm^{\rm max}\simeq 4\pi R^2\Delta r_{_E}W_\pm\simeq 4
\times 10^{56}\,\,\,{\rm s}^{-1}
\label{Nmax}
\end{equation}
                                                                                \noindent
in the vacuum approximation when any suppression of pair creation rate          in comparison with the vacuum value $W_\pm$ given by Eq. (\ref{W1}) 
is neglected. In Eq. (\ref{Nmax}) to get the last equality the field
strength of $5\times 10^{17}$ V cm$^{-1}$ is used.

The luminosity in $e^+e^-$ pairs 
which are created by the Coulomb barrier is $L_\pm^b\simeq\varepsilon_\pm 
\dot N_\pm$, where $\varepsilon_\pm\simeq mc^2+kT_{_{\rm S}}$ 
is the mean energy of created particles and
$\dot N_\pm$ is given by Eq. (\ref{Npm2}).
At $T_{_{\rm S}}\sim 10^{11}$ K, we have $\varepsilon_\pm
\simeq 10^{-5}$ erg and $L_\pm ^b\simeq 3\times 10^{51}$ ergs 
s$^{-1}$. The total (time integrated) energy converted to $e^+e^-$ pairs 
during the first ten seconds after the strange star formation may be as 
high as a few $\times 10^{51}$ ergs (for details, see 
Ref. \cite{Usov97b}). This 
energy coincides with the typical energy output of $\gamma$-ray bursters  
\cite{Fishman} if these enigmatic astronomical objects are cosmological 
in origin \cite{arg} as it was suggested in Ref. \cite{Usov75}.  

Another process of $e^+e^-$ pair creation outside the compact objects 
(neutron and strange stars) is the neutrino-antineutrino annihilation,
$\nu +\bar\nu\rightarrow e^++e^-$ \cite{Goodman87}. The luminosity
of a strange star in $e^+e^-$ pairs which are created in this 
process is \cite{HPA}

\begin{equation}
L_\pm^\nu\simeq 2\times 10^{50}
\left({T_c\over 10^{11}\,{\rm K}}\right)^9
\left({R\over 10^6\,{\rm cm}}\right)^3\,\,\,\,\,{\rm ergs\,\,s}^{-1}\,.
\label{Lnu}
\end{equation}

\noindent
This luminosity may be more than the luminosity of the Coulomb barrier
in $e^+e^-$ pairs only during the first second after 
the strange star formation \cite{ST83} when the temperature $T_c$ 
at the stellar center is higher than $\sim 10^{11}$ K.
Here and below, it is taken into account that strange stars 
with $T_c\lesssim 10^{11}$ K are nearly isothermal (e.g., Ref. 
\cite{HPA}). The value of $L_\pm^\nu$ rapidly 
decreases with the temperature decrease (see Eq. (\ref{Lnu})) while the 
value of $L_\pm^b$ varies quite slowly (roughly, $L_\pm^b\propto T_{_
{\rm S}}\ln T_{_{\rm S}}$ at $10^{10}\lesssim T_{_{\rm S}}\lesssim 10^
{11}$ K). For a strange star with $T_c\ll 10^{11}$ K,
we have $L_\pm^b \gg L_\pm^\nu$, i.e. in this case, $e^+e^-$ pairs
are mainly produced by the Coulomb barrier. 
At $T_{_{\rm S}}\simeq T_c =10^{10}$ K, for example,
from Eqs. (\ref{Npm2}) and (\ref{Lnu}) the ratio $L_\pm^\nu /L_\pm^b$ 
is $\sim 10^{-8}$. As to neutron stars, their luminosity in $e^+e^-$
pairs created due to neutrino-antineutrino annihilation 
\cite{WB92,Goodman87} is not more than the value given by Eq. 
(\ref{Lnu}). Therefore, for a fixed temperature $T_c$ at the stellar 
center, $T_c \ll 10^{11}$ K, we also have $L_\pm^b \gg L_\pm^\nu$,
where $L_\pm^b$ and $L_\pm^\nu$ are the luminosities in $e^+e^-$
pairs for a strange star and a neutron star, respectively. 

Recently, some criteria were suggested \cite{Usov97a} for a compact
X-ray source to be considered as a strange star candidate. 
The considered effect of $e^+e^-$ pair creation by the Coulomb 
barrier may be an additional observational signature of strange 
stars with nearly bare quark surfaces. In connection with this,
it is interesting to note that the emission feature at $\sim 0.5$ MeV 
was observed \cite{Ch} in the spectrum of 1E 1740.7--2942 which answers
the criteria formulated in Ref. \cite{Usov97a} and may be, in fact,
a strange star, not a black hole as it is usually assumed. This feature 
is commonly believed to be related to $e^+e^-$ annihilation into 
two photons. To match the data on this feature, the $e^+e^-$ pair
flux of $\sim 10^{43}$ s$^{-1}$ from the source of pair plasma into 
outer space where pairs annihilate is required. This flux is
equal to the pair flux from a strange star (see Eq. \ref{Npm2})) 
if the surface temperature of the star is $T_{_{\rm S}}\simeq
5\times 10^8$ K. This temperature is consistent with the data 
on the X-ray spectrum of 1E 1740.7--2942 \cite{Ch}.

\end{multicols}

\par\vfill\eject

\noindent
Figure captions

\medskip

\noindent
Fig. 1. Creation of $e^+e^-$ pairs by the Coulomb barrier at the 
quark surface. The electric field of the barrier is very strong,
$E>E_{\rm cr}$, in the region $r_{\rm min}<r<r_{\rm max}$.
Two thick solid lines given by equations Energy = $\pm
mc^2-eV(r)$ restrict the classically forbidden zone. The part of this 
zone where the tunneling of electrons from the "Dirac sea" is
allowed at $T_{_{\rm S}}>0$ is marked by dots.


\begin{references}

\bibitem{W84} E. Witten, Phys. Rev. D {\bf30}, 272 (1984).

\bibitem{AFO86} C. Alcock, E. Farhi, and A. Olinto,
Astrophys. J. {\bf 310}, 261 (1986); P. Haensel, J.L. Zdunik, and
R. Schaeffer, Astron. Astrophys. {\bf 160}, 121 (1986).

\bibitem{G90} N.K. Glendenning, Mod. Phys. Lett. A
{\bf 5}, 2197 (1990); R.R. Caldwell and J.L. Friedman,
Phys. Lett. B {\bf 264}, 143 (1991); J.
Madsen and M.L. Olesen, Phys. Rev. D {\bf 43}, 1069 (1991).

\bibitem{WSWG96} F. Weber, Ch. Schaab, M.K. Weigel, and
N.K. Glendenning, preprint astro-ph/9609067 (1996)

\bibitem{GW92} N.K. Glendenning and F. Weber, Astrophys. J. 
{\bf 400}, 647 (1992)

\bibitem{HPA} P. Haensel, B. Paczy\'nski, and P. Amsterdamski,
Astrophys. J. {\bf 375}, 209 (1991); A. Burrows and J. Hayes, 
Phys. Rev. Lett. {\bf 76}, 352 (1995);
K.S. Cheng and Z.G. Dai, Phys. Rev. Lett. {\bf 77}, 1210 (1996).

\bibitem{WB92} S.E. Woosley and E. Baron, Astrophys. J. {\bf
391}, 228 (1992); A. Levinson and D. Eichler, Astrophys. J. {\bf  
418}, 386 (1993); S.E. Woosley, Astron. Astrophys. {\bf 
97}, 205 (1993).

\bibitem{KWWG} Ch. Kettner, F. Weber, M.K. Weigel, and N.K.
Glendenning, Phys.Rev. D {\bf 51}, 1440 (1995) and references therein.

\bibitem{Usov97a} V.V. Usov, Astrophys. J. Lett. {\bf 481}, L107 (1997).

\bibitem{Schwinger} J. Schwinger, Phys. Rev. {\bf 82}, 664 (1951).

\bibitem{Rein} G. Reinhardt, {\it Quantum Electrodynamics} (Springer,
Berlin, 1994).

\bibitem{Usov97b} V.V. Usov, Astrophys. J. (in preparation).

\bibitem{Gena} G.S. Bisnovatyi-Kogan, {\it Fizicheskie Voprosi Teorii
Zvezdnoi' evolutsii} (Nauka, Moscow, 1989).

\bibitem{Fishman} For a review on $\gamma$-ray bursters see
G.J. Fishman and C.A. Meegan, Ann. Rev. Astron. 
Astrophys. {\bf 33}, 415 (1995); D.H. Hartmann and
S.E. Woosley, Adv. Space Res. {\bf 15}, (5)143, (1995);
C.D. Dermer and T.J. Weiler, Astrophys. Sp. Sci. {\bf 231}, 377 (1995). 

\bibitem{arg} For arguments in favor of cosmological origin
of $\gamma$-ray bursters see M.S. Briggs {\it et al.},
Astrophys. J. {\bf 459}, 40 (1996);
R. Nemiroff {\it et al.}, Astrophys. J. {\bf 435}, 
L133 (1994); J.P. Norris {\it et al.}, Astrophys. J. {\bf 439},
542 (1995); M. Metzger {\it et al.}, IAU Circ. No. 6655 (1997).

\bibitem{Usov75} V.V. Usov and G.V. Chibisov, Sov. Astron. {\bf 19},
115 (1975); S. van den Bergh, Astrophys. Sp. Sci. {\bf 97},
385 (1983).

\bibitem{Goodman87} J. Goodman, A. Dar, and S.  Nussinov,
Astrophys. J. Lett. {\bf 314}, L7 (1987);
D. Eichler, M. Livio, T. Piran, and D. Schramm,
Nature (London) {\bf 340}, 126 (1989).

\bibitem{ST83} A.L. Shapiro and S.A. Teukolsky. {\it Black Holes,
White Dwarfs, and Neutron Stars: Physics of Compact Objects} (Wiley, 
New York, 1983). 

\bibitem{Ch} E. Churazov {\it et al.}, Astrophys. J. {\bf 407}, 
752 (1993) and references therein.

\end{references}
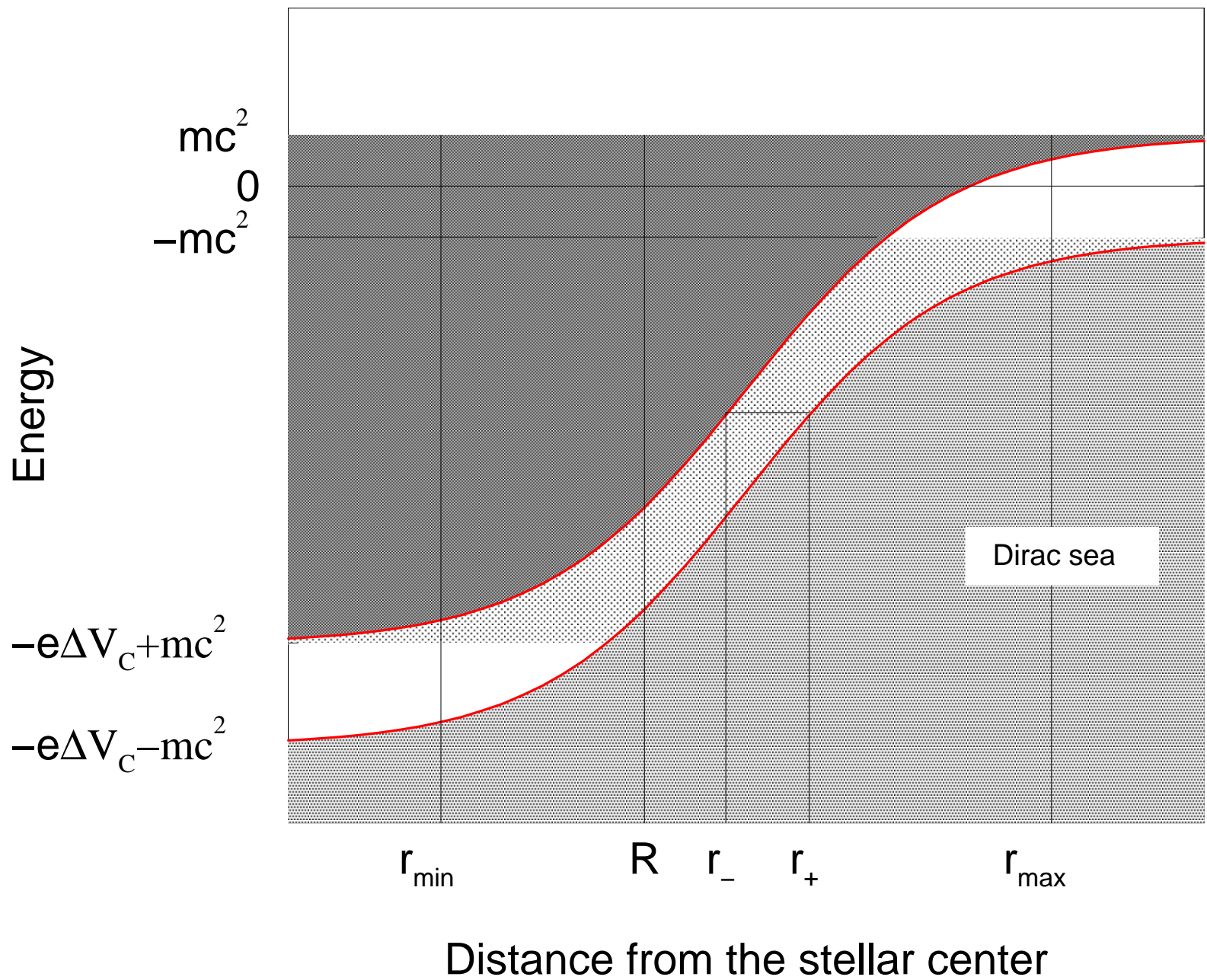
\end{document}